\documentclass{article}

\usepackage{PRIMEarxiv}

\usepackage[utf8]{inputenc} % allow utf-8 input
\usepackage[T1]{fontenc}    % use 8-bit T1 fonts
\usepackage{hyperref}       % hyperlinks
\usepackage{url}            % simple URL typesetting
\usepackage{booktabs}       % professional-quality tables
\usepackage{amsfonts}       % blackboard math symbols
\usepackage{nicefrac}       % compact symbols for 1/2, etc.
\usepackage{microtype}      % microtypography
\usepackage{lipsum}
\usepackage{fancyhdr}       % header
\usepackage{graphicx}       % graphics
\graphicspath{{media/}}     % organize your images and other figures under media/ folder
\usepackage{mwe} 
\usepackage{booktabs}
\usepackage{array}
 
\usepackage{multirow}
\usepackage[normalem]{ulem}
\useunder{\uline}{\ul}{}
\usepackage{float}

\usepackage{amssymb}

\usepackage{graphicx}
\usepackage{amsmath}
\usepackage{booktabs}

\usepackage{color}
\usepackage{comment}
\title{Hierarchical Frequency-based Upsampling and Refining for Compressed Video Quality  Enhancement}
\author{
  Qianyu Zhang \\
  Hangzhou Dianzi University \\
  \texttt{qyzhang@hdu.edu.cn} \\
   \And
  Bolun Zheng \\
  Hangzhou Dianzi University \\
  \texttt{blzheng@hdu.edu.cn} \\
     \And
  Xinying Chen \\
  Hangzhou Dianzi University \\
  \texttt{212320059@hdu.edu.cn} \\
     \And
  Quan Chen \\
  Hangzhou Dianzi University \\
  \texttt{chenquan\_hdu@163.com} \\
       \And
  Zunjie Zhu \\
  Hangzhou Dianzi University, \\
  \texttt{zunjiezhu@hdu.edu.cn} \\
  \And
  Canjin Wang \\
  State Key Laboratory of Media Convergence Production Technology and Systems \& Xinhua Zhiyun Technology Co., Ltd. \\
  \texttt{wangcanjin@shuwen.com} \\
\And
  Zongpeng Li \\
  Hangzhou Dianzi University \\
  \texttt{zongpeng@tsinghua.edu.cn} \\
 \And
  Chenggang Yan \\
  Hangzhou Dianzi University \\
  \texttt{cgyan@hdu.edu.cn} \\
}

\begin{document}
\maketitle

% keywords can be removed

\begin{abstract}

Video compression artifacts arise due to the quantization operation in the frequency domain. 
The goal of video quality enhancement is to reduce compression artifacts and reconstruct a visually-pleasant result.
%Compressed video enhancement needs to accommodate repairing over-smoothed edges while suppressing artifactual edges, such as the ringing effect, for a satisfying result. 
In this work, we propose a hierarchical frequency-based upsampling and refining neural network (HFUR) for compressed video quality enhancement.
HFUR consists of two modules: implicit frequency upsampling module (ImpFreqUp) and hierarchical and iterative refinement module (HIR). ImpFreqUp exploits DCT-domain prior derived through implicit DCT transform, and accurately reconstructs the DCT-domain loss via a coarse-to-fine transfer. Consequently, HIR is introduced to facilitate cross-collaboration and information compensation between the scales, thus further refine the feature maps and promote the visual quality of the final output. 
We demonstrate the effectiveness of the proposed modules via ablation experiments and visualized results.
Extensive experiments on public benchmarks show that HFUR achieves state-of-the-art performance for both constant bit rate and constant QP modes.

\end{abstract}


%\begin{IEEEkeywords}
%Article submission, IEEE, IEEEtran, journal, \LaTeX, paper, template, typesetting.
%\end{IEEEkeywords}

\section{Introduction}
Video compression, {\em a.k.a.} video encoding, is a fundamental technology for transmitting and preserving videos with limited bandwidth and storage.
Video encoding standards, such as H.264/AVC \cite{wiegand2003overview}, H.265/HEVC \cite{sullivan2012overview} and H.266/VVC\cite{bross2021overview}, allow us to encode videos of increasing resolution and growing efficiency. However, compression artifact is inevitably introduced due to quantization and block-based encoding strategies, leading to great loss of fidelity and perceived quality.

\begin{figure}[]
    \centering
    \includegraphics[width=0.5\linewidth]{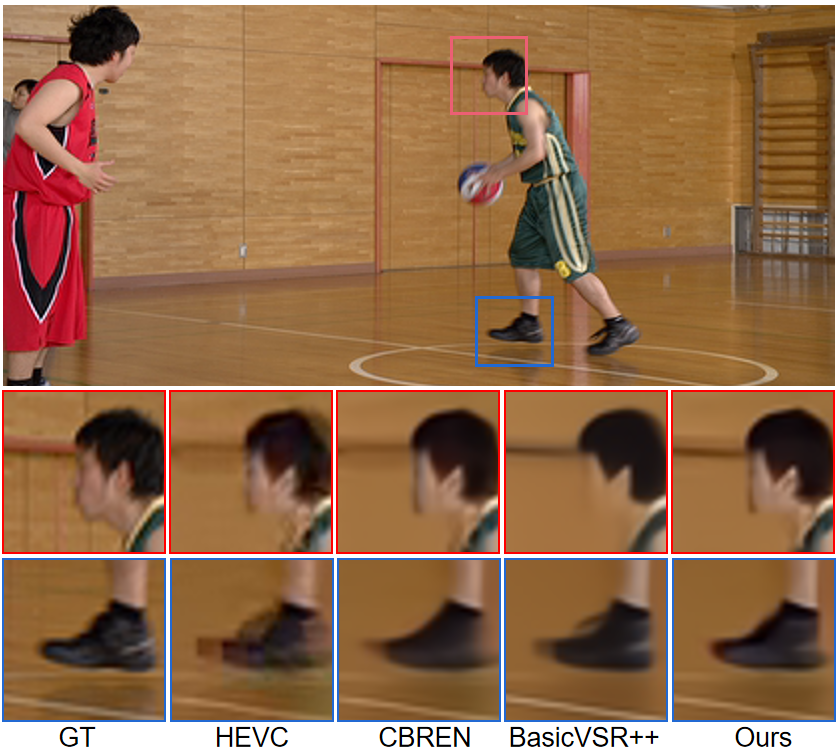}
    \caption{Blurring (blue patch) and artifacts (red patch) in compressed video. Existing methods fall short in reconstructing visually-pleasant outcomes, yielding results that are excessively smooth or still exhibit some artifacts. }
    \label{introduction}
\end{figure}

Substantial efforts have been made in traditional studies %chen2001adaptive,minami1995optimization
\cite{buades2005non,zhang2013compression,sun2007postprocessing,liu2015data,fu2019jpeg} to mitigate the artifact brought by video compression.
However, these approaches suffer from over-smoothed texture details %chen2001adaptive,minami1995optimization,
\cite{buades2005non,zhang2013compression} and prohibitive computational costs for optimization \cite{sun2007postprocessing,liu2015data,fu2019jpeg}. 
Recently, deep neural networks (DNNs) have achieved significant performance improvements in reducing video compression artifact, thanks to their powerful nonlinear modeling capabilities.
These methods can be roughly divided into two categories: pixel-domain methods\cite{jin2020dual,fu2021learning,lee2021wide,zhang2023video,luo2022spatio,zhao2021recursive,deng2020spatio}  and transform-domain methods\cite{chen2018dpw,zheng2019implicit,zhao2021cbren}. %sun2020reduction
Pixel-domain methods focus on the improvement of fusing multiple input frames as well as the enhancement sub-networks, ignoring prior information in the frequency-domain. 
Transform-domain methods seek to reconstruct the transform coefficients to recover high-frequency information with the guidance of prior information of the quantization operation during encoding.
%\cite{sun2020reduction} achieved the removal of compression artifacts by reconstructing the DCT coefficients, \cite{zheng2019implicit} designed an Implicit Dual-domain Convolutional Network directly estimating the DCT-domain losses without DCT to build a color-to-color network. 
%However, in CBR-coded video, bit cost is the primary consideration, and QP changes according to the complexity of the content, thus producing various scales of distortion, which are seldom taken into account by existing methods for compressed video enhancement. 
Video compression technologies employ adaptive quad-tree coding, selecting various coding units (CUs) based on the characteristics of different regions of the image. That produces various scales of distortion, which are rarely taken into account by existing methods for compressed video enhancement. Zhao \textit{et al.}\cite{zhao2021cbren} introduce a multi-scale framework to reduce block distortion of different sizes, yet it fails to compensate for the potential loss of high-frequency information %cross-scale transfer, 
during transmission from high to low scales and tends to produce over-smoothed results. 
%It is essential to exploit high-frequency information for compressed video enhancement. Additionally, higher compression ratios incur severe high-frequency compression artifacts,  and it is pretty challenging to distinguish between high-frequency artifacts and details during cross-scale information transfer. Striking a balance between artifact removal and detail enhancement presents a tremendous challenge.
The process of upsampling represents a unidirectional estimation where the estimated information may not be optimal, thus still leading to some artifacts ({\em e.g.}, jagged contours, blocking effects).
%the balance between smoothing artifact (e.g., jagged contours, blocking effects) and enhancing details is another challenge.
%Exploiting high-frequency details for compressed video enhancement is essential, so focusing on high-frequency information during cross-scale information transfer is desirable. Additionally, higher compression ratios produce severe high-frequency compression artifacts. 
Especially when higher compression rates are applied, it's hard to distinguish between high-frequency artifacts and native details during cross-scale information transfer, which leads to the amplification of produced artifacts or over-smoothed details.

Given the challenges above, we propose HFUR, a DNN-based architecture to hierarchically reconstruct the frequency information via frequency-based upsampling and iterative feature refinement for effective video quality enhancement. 
Specifically, the proposed method is formulated within a multi-scale framework \cite{zhao2021cbren}, to mitigate the distortion of CUs with varying scales. 
An implicit frequency upsampling module (ImpFreqUp) is then introduced to strengthen the cross-scale transfer of frequency information. Prior information brought by the quantization operation is taken into account during the upsampling.
%To address the problem of high-frequency loss during cross-scale information transfer due to the traditional pixel-domain up-sampling, an Implicit Frequency Upsampling (ImpFreqUp) is devised. Building on the observation that the video compression problem arises in the frequency domain, we exploit the DCT domain priors derived by implicit computation during the up-sampling process, introduce an adaptive quantization table, %and increasing the sampling frequency of IDCT to emphasize the preservation of the high-frequency information during the propagation of the high-scale information to the low-scale. 
%and emphasize the preservation of high-frequency information during the propagation of the high-scale information to the low-scale via increasing the sampling frequency of the IDCT.
%Hierarchical Detail Refinement and Artifact Suppression Module 
Furthermore, we design a hierarchical and iterative refinement module (HIR) to refine the feature map upsampled by the ImpFreqUp, aiming at precisely enhancing native details and suppressing produced artifacts.
The HIR roughly separates the features into smooth and sharp components in two branches. 
It optimizes the upsampled features through iterative scale transformations,
hierarchically conducting non-local dependency modeling to suppress artifacts and local adjustments to enhance details.
Through extensive experiments on public benchmarks, we demonstrate the effectiveness of ImpFreqUp and HIR, and the superiority of our hierarchical frequency-based upsampling and refining neural network.
%Since the image is divided into small patches during compression and treated independently of each other, which tends to produce artifacts or distortions in specific areas of the image rather than globally, FIRM separately undertakes local detail adaptation and modeling long-distance dependence, facilitating the cross-collaboration and information complementarity across scales through hierarchical refinement. It strikes a balance between the elimination of artifacts and the enhancement of details. 
Generally, our contributions are summarized as follows:
\begin{itemize}
	%\item We exploit the DCT domain priors derived through implicit computation in the upsampling process to attend to high-frequency information during cross-scale information transfer. An Implicit Frequency Upsampling is proposed by increasing the sampling frequency of the IDCT to guarantee the retention of high-frequency information as it propagates from high to low scales. 
    \item We propose a novel frequency-based upsampling method called ImpFreqUp via implicit DCT transform to accurately reconstruct frequency information during cross-scale transfer.
    \item We design a hierarchical and iterative refinement module that
    separates the input into two complementary features at different scales,
hierarchically facilitating cross-collaboration and information compensation between scales to further refine the feature map produced by the ImpFreqUp.

   % separates the input into two complementary features at different scales,  conducting local details adjustment and non-local dependency modeling in parallel,
   % hierarchically facilitating cross-collaboration and information compensation between scales to further refine the feature map.
   % to strike a  balance between enhancing the native details and suppressing the produced artifacts.
    \item We design a hierarchical frequency-based upsampling and refining neural network namely HFUR for compressed video quality enhancement. In performance evaluation of video compression enhancement, our HFUR achieves state-of-the-art results.
    
    %stands out from the mise-en-scene of its state-of-the-art alternatives. 
    %by effectively suppress distortion, ringing, and artifacts in compressed video.
	%\item We conduct extensive experiments to evaluate the proposed approach, and our method accomplishes state-of-the-art video compression enhancement.
 
 %We design a  XXX network to deal with quality instability due to different QPs under CBR.The proposed model achieves state-of-the-art performance under both CBR and CQP.

\end{itemize}

\section{Related Work}
\textbf{Pixel-domain based methods.}
The past decade has witnessed substantial developments in pixel-domain compressed video enhancement. ARCNN\cite{dong2015compression} first proposes a deep learning scheme consisting of four convolution layers, to train the mapping function from the compressed image to the reconstructed image. Tai \textit{et al.} \cite{tai2017memnet} introduce LSTM to image restoration, and propose a deep memory network using recursive and threshold units to construct a memory module. Galter \textit{et al.}\cite{galteri2017deep} adopt generative adversarial networks and employ structural similarity loss instead of mean square error loss to generate more realistic image details for better visual sensory effects in reconstructed images. Jin \textit{et al.} \cite{jin2020dual} propose dual-stream recurrent networks to deal with specific artifacts in high-frequency and low-frequency components respectively, and to reduce the overall number of parameters of the network through a parameter-sharing mechanism. Fu \textit{et al.}\cite{fu2021learning} increase the interpretability of the artifact removal network by respectively extracting pixel-level and semantic-level features, modeling and solving pixel-level prior and semantic-level prior so that the network obtains better artifact removal performance.

These approaches consider single-frame information only, and ignore information in the temporal domain.  %guan2019mfqe
MFQE\cite{yang2018multi} proposes a lightweight multi-frame framework exploiting Peak Quality Frames to enhance other low-quality frames. STDF\cite{deng2020spatio} employs spatio-temporal deformable convolution to aggregate temporal information to reduce the effect of inaccurate optical flow. Based on STDF, RFDA\cite{zhao2021recursive} proposes a recursive fusion module to model temporal dependencies over a long period. TSAN\cite{xu2022transcoded} aims at transcoding video recovery, and uses temporal deformable alignment and pyramidal space fusion to tackle it. BasicVSR++\cite{chan2022basicvsr++} uses spatio-temporal information more effectively across mismatched video frames by presenting second-order lattice propagation and flow-guided deformable alignment. STCF\cite{zhang2023video} proposes a CNN-Transformer-based framework to exploit the global information modeling adequately. Although existing approaches evolved in the pixel domain for video enhancement, the video compression problem arises in the frequency domain, and thus, several approaches are explored in the frequency domain.

\textbf{Frequency-domain methods.}
Numerous researches have investigated learning in the frequency domain, both high-level semantic tasks  %ehrlich2019deep
\cite{xu2020learning,qin2021fcanet} and low-level restoration tasks \cite{wang2016d3,ehrlich2020quantization}.
Several low-level approaches have explored the restoration of content details from the perspective of frequency decomposition. Li \textit{et al.}\cite{li2021learning} decompose features into different frequency bands via multi-branch CNN. Other studies \cite{ehrlich2020quantization,wang2016d3} have converted images to the frequency domain. For example, Chen \textit{et al.}\cite{chen2018dpw} propose a discrete wavelet transform-based method to map compressed images from pixel domain to DWT domain, exploiting soft decoding to improve image quality without introducing additional coding bits. Recently, the discrete cosine transform (DCT) domain has been introduced for frequency analysis. Frequency application as introduced in CNNs via JPEG coding \cite{xu2020learning,gao2021neural}. %,liu2018frequency}. 
Guo \textit{et al.}\cite{guo2016building} jointly learn a deep convolutional network in both DCT and pixel domains, helping leverage the prior knowledge of DCT in the JPEG compression domain. Wang \textit{et al.}\cite{wang2016d3} design a dual-domain restoration network for removing artifacts from JPEG-compressed images. In addition, Ehrlich \textit{et al.}\cite{ehrlich2020quantization} devise a y-channel correction network and a color-channel correction network to correct JPEG artifacts. FTVSR \cite{qiu2022learning} conducts self-attention over a joint space-time-frequency domain to recover the high-frequency details. IDCN\cite{zheng2019implicit} proposes an implicit dual-domain convolutional network that implicitly exploits pixel-domain features and DCT-domain prior. CBREN \cite{zhao2021cbren} designs a multi-scale framework to reduce block distortion at different scales of compressed video, but like most multi-scale frameworks, it employs regular pixel-domain upsampling to recover the resolution without exploiting prior information in the frequency domain.
%but The multi-scale framework is common in image/video enhancement tasks for dealing with multi-scale compression degradation. However, these approaches employ normal pixel-domain up-sampling to recover the resolution without exploiting a priori information in the frequency domain.

\textbf{Upsampling}
Upsampling plays a critical role in multi-scale modeling, such as feature pyramids \cite{lin2017feature,seferbekov2018feature} and image pyramids \cite{luo2020lightweight,liu2020ipg}, where it is utilized to increase the resolution of the image and obtain more detailed information. The most common methods, such as\cite{jing2009improved,keys1981cubic,reichenbach2003two,dengwen2010edge}, arrive at the value of the target pixel through the values of spatially neighboring pixels. Due to the fixed up-sampling filter, the high-frequency portion of the reconstructed image tends to produce annoying artifacts such as blocking, edge jaggedness, and ringing effects. Recently, \cite{shi2016real} has proposed sub-pixel convolutional layer, which efficiently and flexibly implements up-sampling, and has been widely used in image reconstruction. However, most existing work serves up-sampling in the spatial domain and rarely explores the potential of up-sampling in the frequency domain. %\cite{} Zero appending in high-frequency components usually leads to blocking artifacts.

\begin{figure}[htbp]
    \centering
    \includegraphics[width=0.8\linewidth]{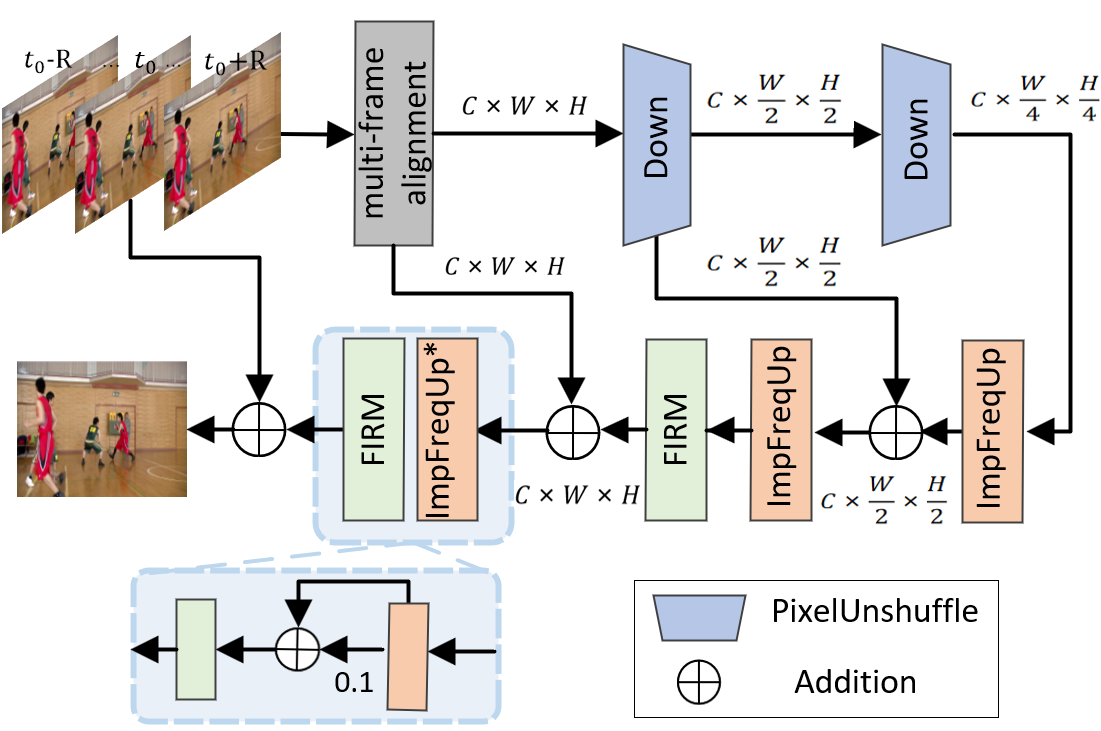}
    \caption{Overview of HFUR for compressed video quality enhancement. HFUR consists of two modules: implicit frequency upsampling module and hierarchical and iterative refinement module. Note that ImpFreqUp$^*$ is a special upsampling module that achieves $\times 1$ upsampling. }
    \label{model}
\end{figure}

\section{Methods}

In this section, we explore frequency domain prior for compressed video, and design a hierarchical frequency-based upsampling and refining neural network. Since HEVC adopts inter-frame compression to reduce temporal redundancy, leveraging temporal sequence information for motion compensation is also critical. Many studies \cite{deng2020spatio,zhao2021recursive,luo2022spatio,wang2019edvr,chan2022basicvsr++} have produced significant achievements on the spatio-temporal feature fusion module. Following \cite{wang2019edvr,zhao2021cbren}, we adopt the PCD alignment module\cite{wang2019edvr} for multi-frame alignment.
To reduce compression artifacts and reconstruct a visually-pleasant result, ImpFreqUp is introduced to accurately reconstructs the DCT-domain loss via a coarse-to-fine transfer. Then, HIR is proposed to facilitate cross-collaboration and information compensation between the 
scales. The overall architecture of the framework is shown in Fig.~\ref{model}.

\subsection{Preliminary}
Compression artifacts arise from the quantization of the DCT coefficient matrix. Let's assume $\Theta$ is the coefficient matrix and $\Theta ^{\ast}$ is the quantized version, the quantization loss $\xi$ of a compressed video can be expressed as:
{\small
\begin{equation}
   \xi = \Theta - \Theta ^{\ast}
   \label{eq1}
\end{equation}
}

The pixel domain loss can be expressed as:
{\small
\begin{equation}
  L_{p}=T^{-1}_{DCT} ( \xi ) = T^{-1}_{DCT} ( \Theta - \Theta ^{\ast} )
   \label{eq2}
\end{equation}
}
where $T^{-1}_{DCT}$ denotes the inverse DCT.
Since $T^{-1}_{DCT}$ is a linear transformation, the video compression distortion can be expressed as a residual structure. Most existing methods use CNN \cite{zhao2021cbren,zhao2021recursive} or transformer \cite{zhang2023video} with residual structure as a baseline, essentially aiming to estimate feature-domain representations of compression distortion. Consequently, we consider the estimated quantization loss in the feature domain as:
{\small
\begin{equation}
  L_{f}= Conv(L_{p}) = Conv (T^{-1}_{DCT} ( \xi ))
   \label{eq22}
\end{equation}
}
where $ Conv$ denotes a convolution or transformer based operation.
%achieving the transformation from the pixel domain loss $L_{p}$ to the feature domain loss $L_{f}$. 
As illustrated in Eq.~\ref{eq22}, the key to estimate $L_{f}$ lies in the precise estimation of $\xi$.
Inspired by \cite{zheng2019implicit}, we can introduce a set of convolutions to directly estimate the $\xi$ without explicit supervision. 
Since $\xi$ is generated by the quantization of HEVC, we further deconstruct it into $\delta$ and $T^{qp}$:
{\small
\begin{equation}
     \xi = \delta \  {\ast} \  T^{qp}
     \label{eq3}
\end{equation}
}
where $\ast$ is element-wise multiplication, $T^{qp}$ is a $p \times p$ quantization table under the specified quantization parameter (QP), and $\delta$ is relative quantization loss, which is a $p\times p$ matrix and restricts:
{\small
\begin{equation}
    -0.5 < \delta_i < 0.5 \ \ \  \  \forall{\delta_i \in \delta}
    \label{eq4}
\end{equation}
}  
Therefore, we can separately estimate the $\delta$ and $T^{qp}$ to leverage the prior information, instead of estimating $\xi$ directly. 

%Similar to traditional method to calculate the artifacts in compressed video, we use convolution layer to estimate $\delta$. According to Eq.~\ref{eq4}, we introduce two constraints of ${\delta}_i$:
%{\small
%\begin{equation}
%\Gamma(\delta)=\left\{\begin{matrix} 
%-0.5 & \delta_i < -0.5\\  
%\delta_i & -0.5 < \delta_i < 0.5\\
%0.5 & \delta_i > 0.5
%\end{matrix}\right.
%\end{equation}
%}  

%While there is no supervision to explicitly ensure the accuracy of the estimated DCT coefficient loss, back-propagation during training will drive the neural network to achieve roughly favourable estimation results.

\subsection{Implicit Frequency Upsampling}
HEVC describes an extensive range of block sizes up to $64\times 64$ pixels, with adaptive quad-tree coding using a coding tree unit. To mitigate block distortions with various scales, we adopt a multi-scale structure as shown in Fig.~\ref{model}.
Existing multi-scale upsampling components \cite{zhao2021cbren,zhang2022multi} generally derive the target output by mixing neighboring elements of the feature domain. However, such upsampling methods struggle to focus on high-frequency information during cross-scale transfer, and tends to produce over-smoothed results \cite{schwarz2021frequency}. In this case, careful consideration should be given to maximizing information transfer from higher to lower scales, and making a more precise estimation at lower scales. Encouraged by the fact that the video compression artifacts arise from quantization in the DCT domain, we naturally introduce a DCT-domain prior in the upsampling process and accurately reconstruct the DCT-domain loss. %via \textcolor{red}{a coarse-to-fine transfer.}
\begin{figure}[htp]
    \centering
    \includegraphics[width=0.8\linewidth]{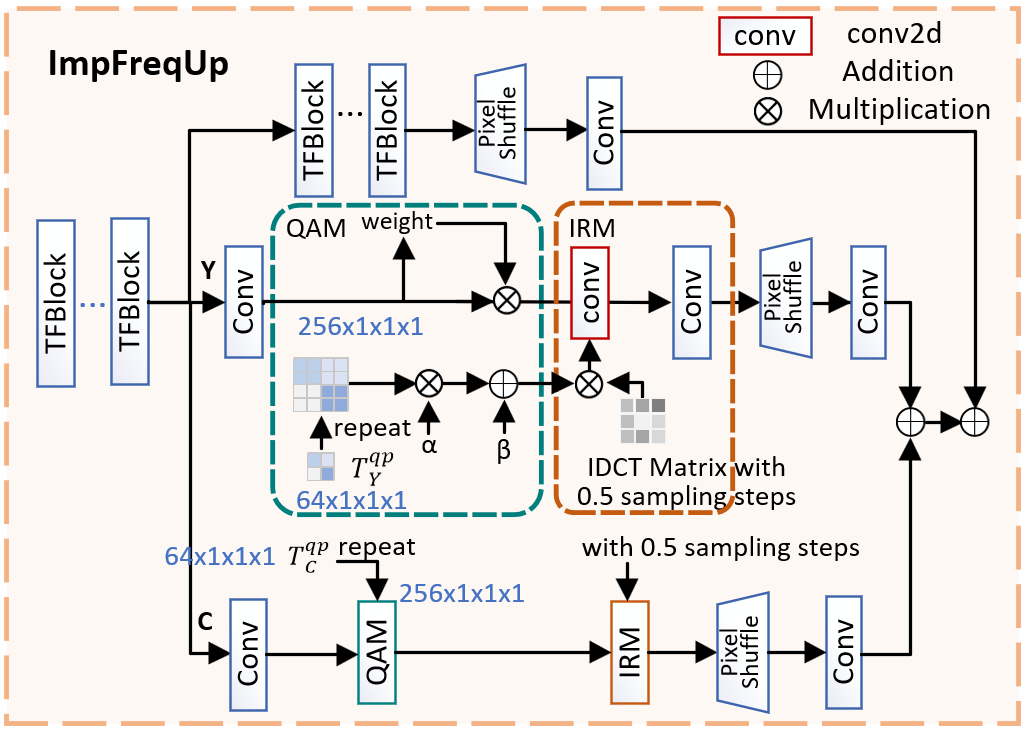}
    \caption{Architecture of ImpFreqUp.}
    \label{ImpFreqUp}
\end{figure}

Given an image patch $P$ with the size of $N_p \times N_p$, the $P_i$ denotes a $\frac{N_p}{2^i} \times \frac{N_p}{2^i}$ sized patch acquired by $P$ at the $i$-th ($i \in \{1,2,3\}$) scale. 
To approximate $P_i$, we design a DCT domain approach as: 
{\small
\begin{equation}
 argmin_{\hat{\Theta}_i}(T_{DCT}^{-1}(\hat{\Theta}_i)-P_i)
\end{equation}
}
where the $\hat{\Theta}_i$ denotes the estimated DCT coefficient matrix for $P_i$ and $T^{-1}_{DCT}$ denotes the inverse DCT function. 
For multi-scale network structures, we introduce an implicit upsampling:
\begin{equation}
 argmin_{\Theta_{i+1}}(S(\hat{\Theta}_{i+1})-P_i)
\end{equation}
where $S$ denotes the upsampling function. According to Eq.~\ref{eq1}, we can achieve $\Theta$ via estimating $\xi$ with residual structures.%rather than directly estimate $\Theta$. 
Therefore, based on Eq.~\ref{eq2}, we can use IDCT to reconstruct the spatial signal with $\xi$ as: %the DCT coefficient matrix $\xi$ as:
{\small
\begin{equation}
T_{DCT}^{-1}(\xi)_{x,y} = \sum_{u=0}^{N_p-1} \sum_{v=0}^{N_p-1} \alpha(u)\alpha(v)f(x,y,u,v)
\end{equation}
\begin{equation}
f(x,y,u,v) = \xi(u,v)cos(\frac{2x+1}{2N_p}u\pi)cos(\frac{2y+1}{2N_p}v\pi)
\end{equation}
}
where $x$ and $y$ denote the horizontal and vertical coordinates in the $N_{p}\times N_{p}$ image patch, $\alpha(\cdot)$ is a coefficient function that can be written as:
{\small
\begin{equation}
\alpha (u)=\left\{\begin{matrix} 
  \sqrt{ \frac{1}{N_p} } \ \ \ \ \ \  u=0\\  
  \sqrt{ \frac{2}{N_p} } \ \ \ \ \ \  u> 0
\end{matrix}\right. 
\end{equation}
}
Noticing that the classic IDCT is a cross-domain
sampling function, we accomplish the upsampling with $\xi$ by
expanding the sampling rate of the IDCT function to achieve
$\times 2$ upsampling:
{\small
\begin{equation}
\begin{split} 
S_{x,y} = \{T_{x-0.25,y-0.25}^{-1},T_{x-0.25,y+0.25}^{-1},\\
T_{x+0.25,y-0.25}^{-1},T_{x+0.25,y+0.25}^{-1}\}
\label{eq9}
\end{split}
\end{equation}
}
\begin{figure*}[thp]
    \centering
    \includegraphics[width=1.0\linewidth]{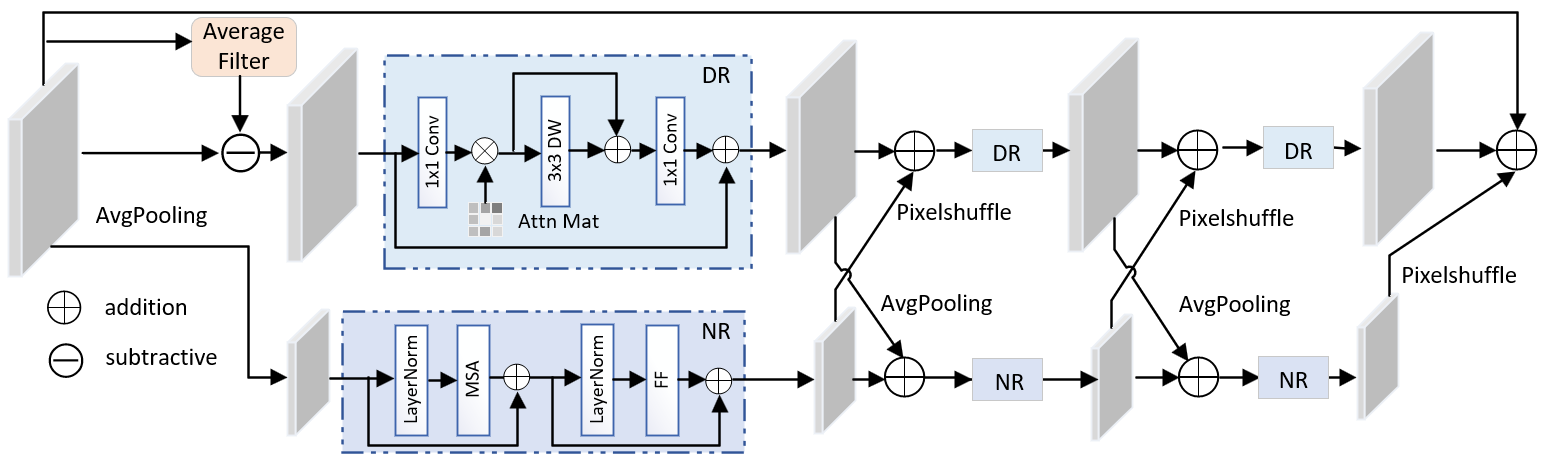}
    \caption{Architecture of HIR}
    \label{FIRM}
\end{figure*}
Fig.~\ref{ImpFreqUp} illustrates the details of the proposed implicit frequency upsampling module (ImpFreqUp). First, we extract feature $F$ with $N_1$ transformer blocks to obtain a large receptive field. Then, a pixel-domain restoration branch and a DCT-domain restoration branch are introduced in parallel, to compute the artifactual representations from the pixel and DCT domains respectively. For the pixel-domain restoration branch, we estimate the pixel-domain loss directly from the input features $F$ by $N_2$ transformer blocks.
In the DCT domain restoration branch, we estimate the loss following Eq.~\ref{eq22} and Eq.~\ref{eq3} for luminance and chrominance channels respectively. Initially, $\delta$ is estimated by a $3 \times 3$ convolutional layer, constrained by Eq.~\ref{eq4}.
Then we adopt an implicit reconstruction module (denoted as IRM in the Fig.~\ref{ImpFreqUp}) \cite{zheng2019implicit}. The sampling interval of conventional IDCT is 1. According to Eq.~\ref{eq9}, we improve the interval to sampling at 0.5 steps to increase the sampling points, to estimate more high-frequency information in the pixel domain accurately.
HEVC supports four transform block sizes: $4\times 4$, $8\times 8$, $16\times 16$, and $32\times 32$. Given the relatively minor distortion caused by $4\times 4$ transform blocks, we set the basic processing size of ImpFreqUp as $8 \times 8$, and obtain $8 \times 8 \times16\times 16$ transform matrix. Then, the matrix is reshaped into a $1 \times 1 \times 64 \times 256$ vector, so we would apply a simple convolution to simulate the IDCT process. 

To estimate $\xi$ in Eq.~\ref{eq3}, we design a quantization aware module (denoted as QAM in the Fig.~\ref{ImpFreqUp}).
In order to match the shape of IDCT matrix, we upsample $8 \times 8 $ sized $T_{base}^{qp}(u,v)$ to get $16 \times 16 $ sized $T_{up}^{qp}(u,v)$ and resize it to $256 \times 1 \times 1 \times 1$. Specifically, the interpolated pixels are the same as the original pixels in a $2 \times 2$ localized region:
{\small
\begin{equation}
\begin{split}
T_{up}(2u,2v) = T_{up}(2u + 1, 2v) = T_{up}(2u, 2v + 1)\\
= T_{up}(2u + 1, 2v + 1) = T_{base}(u, v)
\end{split}
\end{equation}
}
where $u,v\in\{0, 1, . . . , 7\}$ and $T_{base}$ is the basic quantization table defined in HEVC. Note that the quantization matrix for the luminance branch is different from that for the chrominance branch.
The compressed video in constant bit rate coding has different quantization parameters at different positions in the same frame. We introduce two learnable matrices $\alpha$, $\beta$ of dimension $256 \times 1 \times 1$, and conduct adaptive estimation of $T_{qp}$ by affine transformation:
{\small
\begin{equation}
T_{qp} = \alpha · T_{base} + \beta
\end{equation}
}
Specifically, for $\times 1$ ImpFreqUp (denoted as ImpFreqUp* in Fig.~\ref{model}), the sampling interval in IRM is set to 1. 
Thus we get the $8 \times 8 \times8\times 8$ transform matrix which would be reshaped into a $1 \times 1 \times 64 \times 64$ vector, and set the sizes of learnable matrices $\alpha$, $\beta$ to $64 \times 1 \times 1$ to match the original $T_{base}^{qp}$. 
Moreover, the pixelshuffle layers in ImpFreqUp would also be removed as there is no need for scale transformation.

\subsection{Hierarchical and Iterative Refinement}

%In the video compression process, a higher compression ratio not only leads to the loss of a large amount of high-frequency information, but also to inaccurate reconstruction of low-frequency information, therefore, it is desirable to calibrate the low-frequency information and enhance the detail information.

%Though the ImpFreqUp pays attention to more high-frequency information, facilitating the restoration of some of the lost detail to a certain extent, such detail is usually estimated. 
ImpFreqUp reconstructs the DCT-domain loss via a coarse-to-fine transfer, achieving sharper edges. However, the loss is estimated and is incapable of addressing certain generated artifacts, such as jagged contours, blocking effects, and color bleeding. Conventional methods lead to the loss of high-frequency details while suppressing artifacts, which result in blurring effects. 
Therefore, we design the hierarchical and iterative refinement module to facilitate cross-collaboration for better estimation, further refine the feature maps across different scales and optimize the visual quality of the final output.

As shown in Fig.~\ref{FIRM}, we approximately extract the high frequency and low frequency branches from the feature map, obtaining two complementary features at different scales.
Given a feature $F_{in}$, we receive the initial high frequency details $D_f$ from the following equations:
{\small
\begin{equation}
D_{f} = F_{in} - Avg(F_{in})
\end{equation}
}
where $Avg(\cdot )$ denotes the $3 \times 3$ average filtering. Then, we perform localized detail adjustments via the detail refinement module(denoted as DR in Fig.~\ref{FIRM}), formulated as:
{\small
\begin{equation}
\begin{split} 
& F' = D_{f}^{in} + SA( Conv_{1\times 1}(D_{f}^{in}))\\
& F'' =Conv_{1\times 1}( F' + DWConv_{3 \times 3}(F') )\\
& D_{f}^{out} = D_{f}^{in} + F''
\end{split}
\end{equation}
}
where $SA(\cdot )$ denotes self-attention\cite{vaswani2017attention}.
Artifacts arising from video compression are often localized, manifested by distortions in specific regions of the image rather than affecting the entire image. we take advantage of this and introduce the low frequency branch via downsample the $F_{in}$ to one-half the original resolution and enlarge network receptive fields while simultaneously reducing computations:
{\small
\begin{equation}
L_{f} = AvgPool(F_{in})
\end{equation}
}
Then, we introduce the non-local refinement module (denoted as NR in Fig.~\ref{FIRM}), which aims to consider a broader context rather than focusing on minute details, thereby mitigating the impact of local artifacts to some extent.
To leverage the information from these two branches at different scales, enabling synergistic enhancement of high and low-frequency information, we downsample $D_{f}$ via average pooling to complement $L_{f}$ in the low frequency branch and upsample $L_{f}$ via PixelShuffle to complement $D_{f}$ in the high-frequency branch.
Thus we can realize the cross-collaboration between high-frequency and low-frequency features, which not only promotes information complementarity, but also establishes cross-residual linkages for better feature propagation.
%\begin{equation}
%D_{f} = D_{f} + L_{f}\uparrow 
%\end{equation}
%\begin{equation}
%L_{f} = L_{f} + D_{f}\downarrow 
%\end{equation}

\section{Experiments}

\subsection{Dataset}
%\subsubsection{NTIRE 2021 Dataset}
We adopt the dataset proposed in NTIRE2021 quality enhancement of heavily compressed video challenge\cite{yang2021ntire1}, to training our models.
This dataset contains 200 videos that 10 representative videos are selected for validation during the training stage, while the remaining 190 videos serve as the training set. 
For testing, we use 18 standard test sequences from the Joint Collaborative Team on Video Coding (JCT-VC) database. These video sequences cover various resolutions, including Class A (2560 $\times$ 1600), Class B (1920 $\times$ 1080), Class C (832 $\times$ 480), Class D (480 $\times$ 240), and Class E (1280 $\times$ 720).
We conducted experiments on both constant bit rate (CBR) and constant QP (CQP) modes with these data sets. 
In CQP mode, all video sequences are compressed by HM 16.20 with HEVC LowDelay-P (LDP) configuration. To evaluate performance under different compression levels, the compression is conducted with QPs of 27 and 37.
In CBR mode, We adopt settings from recent literature \cite{zhao2021cbren}, the videos would be encoded by libx265-supported FFmpeg at a fixed base bit rates of 200kbps and 800kbps, since the compressed video quality in CBR mode is related to the bit rate. 
%We use 18 HEVC standard test sequences as test dataset, and 
We set different bit rates according to the information of the LDV official documents for different test sets as follows:
{\small
\begin{equation}
Test_{bit} = \frac{Test_{rate}\times Test_{w} \times Test_{h}}{30 \times \ 960 \times 536} \times Base_{bit}
\end{equation}}
\begin{table}[]
  \centering
  \caption{Overall performance comparison of $\Delta$PSNR (dB) over the test sequences at CBR mode. %Video resolution: Class A (2560 $\times$ 1600), Class B (1920 $\times$ 1080), Class C (832 $\times$ 480), Class D (480 $\times$ 240), Class E (1280 $\times$ 720)
  }
% \resizebox{\linewidth}{!}{
    \begin{tabular}{c|cccccc}
    \toprule
    \toprule
    \multicolumn{7}{c}{800kbps} \\
    \midrule
    Method & A     & B     & C     & D     & E     & Average \\
    \midrule
    IDCN  & 0.51  & 0.39  & 0.69  & 0.67  & -0.12 & 0.43 \\
    EDVR  & 0.50  & 0.23  & 0.61  & 0.68  & \underline{0.34}  & 0.47 \\
    STDF  & 0.45  & 0.23  & 0.52  & 0.49  & 0.11  & 0.36 \\
    MIRNet & 0.52  & \underline{0.52}  & 0.68  & 0.67  & 0.20  & 0.52 \\
    CBREN & \underline{0.73}  & 0.49  & \underline{0.85}  & \underline{0.88}  & 0.32  & \underline{0.65} \\
    BasicVSR++ & 0.64  & 0.39  & 0.82  & 0.82  & \textbf{0.35}  & 0.60 \\
    HFUR(ours)  & \textbf{0.89}  & \textbf{0.55}  & \textbf{1.06}  & \textbf{1.11}  & 0.31  & \textbf{0.78} \\
    \midrule
    \multicolumn{7}{c}{200kbps} \\
    \midrule
    IDCN  & 0.31  & 0.42  & 0.37  & 0.23  & 0.43  & 0.35 \\
    EDVR  & 0.46  & 0.38  & 0.54  & 0.43  & -0.27 & 0.31 \\
    STDF  & 0.35  & 0.25  & 0.48  & 0.37  & -0.22 & 0.25 \\
    MIRNet & 0.34  & 0.44  & 0.44  & 0.28  & \textbf{0.52}  & 0.40 \\
    CBREN & \underline{0.64}  & \underline{0.53}  & \underline{0.72}  & \underline{0.56}  & \underline{0.48}  & \underline{0.59} \\
    BasicVSR++ & 0.60  & 0.39  & 0.61  & 0.40  & 0.13  & 0.42 \\
    HFUR(ours)  & \textbf{0.84}  & \textbf{0.64}  & \textbf{0.88}  & \textbf{0.67}  & 0.28  & \textbf{0.66} \\
    \bottomrule
    \bottomrule
    \end{tabular}%
 %   }
  \label{exp_CBR}%
\end{table}%

\subsection{Implementation Details}

%The proposed FRUIT has 4 transformer blocks to extract featureand 4 transformer blocks estimate the pixel-domain loss. Then we use 3 XXX........
In the proposed HFUR, we set the basic processing scale of ImpFreqUp to $8\times 8$, and use $\times 1 \times 2 \times 4$ multi-scale schemes to address distortions at different scales introduced by HEVC as the max transform block size is $32\times32$.
In each ImpFreqUp, the number of TFBlocks is specified as $N_1=4$ and $N_2=4$.
All trainable convolution layers have 64 channels.
For HIR, the input with 64 channels is divided into two branches, each consisting of 32 channels.

We use five consecutive video frames as input.
The training samples are randomly cropped from raw and the corresponding compressed video frames with the size of $96 \times 96$. 
The 8 training samples augmented by random rotation and flipping formulate a training batch.
The Cosine Annealing scheme \cite{loshchilov2016sgdr} and Adam
optimizer \cite{kingma2014adam} with $\beta_1=0.9$ and $\beta_2=0.999$ are used to train our model, while the learning rate is initialized as $4\times 10^{-4}$. We initialize deeper networks by parameters from shallower ones for faster convergence. 
The charbonnier penalty function\cite{lai2017deep} is adopted as the final loss to optimize the model.
We use the Pytorch framework for our implementation, and train on an Nvidia RTX 3090 GPU.

\subsection{Comparison with State-of-the-Art Approaches}
\begin{table}[!bp]
  \centering
  \caption{Overall performance comparison of $\Delta$PSNR (dB) over the test sequences at constant QP mode.}
% \resizebox{\linewidth}{!}{
    \begin{tabular}{c|cccccc}
    \toprule
    \toprule
    \multicolumn{7}{c}{QP37} \\
    \midrule
    Method & A     & B     & C     & D     & E     & Average \\
    \midrule
    IDCN  & 0.56  & 0.47  & 0.71  & 0.67  & 0.76  & 0.63 \\
    EDVR  & 0.64  & 0.55  & 0.79  & 0.80  & 0.82  & 0.72 \\
    STDF  & 0.53  & 0.43  & 0.59  & 0.59  & 0.69  & 0.56 \\
    MIRNet & 0.67  & 0.53  & 0.76  & 0.71  & 0.84  & 0.69 \\
    CBREN & 0.73  & 0.58  & 0.83  & 0.87  & 0.87  & 0.77 \\
    BasicVSR++ & \underline{0.92}   & \underline{0.69}   & \underline{0.96}    & \underline{1.01}    & 0.86  & \underline{0.88} \\
    STCF & 0.85  & 0.68  & 0.89    & 0.94  & \underline{0.93}    & 0.85\\
    HFUR(ours)  & \textbf{1.01}  & \textbf{0.82}  & \textbf{1.12}  & \textbf{1.18}  & \textbf{0.95}  & \textbf{1.01} \\
    \midrule
    \multicolumn{7}{c}{QP27} \\
    \midrule
    IDCN  & 0.54  & 0.40  & 0.71  & 0.76  & 0.57  & 0.60 \\
    EDVR  & 0.64  & 0.49  & 0.87  & 1.06  & 0.64  & 0.74 \\
    STDF  & 0.47  & 0.32  & 0.54  & 0.64  & 0.48  & 0.49 \\
    MIRNet & 0.63  & 0.47  & 0.79  & 0.84  & 0.61  & 0.67 \\
    CBREN & 0.69  & 0.51  & 0.91  & 1.10  & \underline{0.68}  & 0.78 \\
    BasicVSR++ & \underline{0.86} & \underline{0.66} & 1.02   & \underline{1.36}   & \underline{0.68}   &  \underline{0.92} \\
    STCF & 0.83   & 0.63   & \underline{1.03}    & 1.32     & \textbf{0.80}      &  \underline{0.92}     \\
    HFUR(ours)  & \textbf{1.01}  & \textbf{0.79}  & \textbf{1.28}  & \textbf{1.55}  & \textbf{0.80}  & \textbf{1.09} \\
    \bottomrule
    \bottomrule
    \end{tabular}%
%    }
  \label{exp_CQP}%
\end{table}%
In this section, we compare our HFUR with several
state-of-the-art approaches, including IDCN \cite{zheng2019implicit},
EDVR\cite{wang2019edvr}, BasicVSR++ \cite{chan2022basicvsr++}, MIRNet\cite{zamir2020learning}, STDF\cite{deng2020spatio},  
CBREN\cite{zhao2021cbren}, STCF\cite{zhang2023video}. Among them, IDCN and MIRNet are designed for single compressed image enhancement, both EDVR and CBREN utilize the PCD module to achieve alignment which is the same as ours, while the STDF and STCF adopt another alignment and fusion strategy. 
It should be also noticed that the CBREN is specially designed for CBR compressed videos, the STDF and STCF are originally proposed for CQP compressed videos, while the BasicVSR++ is designed for compressed video super-resolution.%leveraging information from the entire input video.
For a fair comparison, we use the official codes retrained on the same dataset and under the same experimental settings.
All compared methods adopt five consecutive frames as input if they allow.
\begin{table}[t]
  \centering
  \caption{Averaged SD for $\Delta$PSNR measured on the class B at QP=27, 32 and BR=200kbps, 800kbps.}
    % \resizebox{\linewidth}{!}{
    \begin{tabular}{c|cccc}
    \toprule
    \toprule
    Method & QP27  & QP37  & 200kbps & 800kbps \\
    \midrule
    HEVC  & 0.74  & 0.92  & 0.85      & 0.67 \\
    IDCN  & 0.71      & 0.92  & 0.85      & \textbf{0.66} \\
    EDVR  & 0.67      & 0.88  & 0.84      & 0.67 \\
    STDF  & 0.85      & 0.88  & 0.85      & 0.68 \\
    MIRNet & 0.70      & 0.91  & 0.86      & 0.67 \\
    CBREN & 0.65      & 0.84  & 0.81      & 0.87 \\
    BasicVSR++ & 0.65      & 0.79  & 0.80      & \textbf{0.66} \\
    HFUR(ours)  & \textbf{0.61}  & \textbf{0.77}  & \textbf{0.78}   & \textbf{0.66} \\
    \bottomrule
    \bottomrule
    \end{tabular}%
 %   }
  \label{tab:sd}%
\end{table}%
We use $\Delta$PSNR as the objective evaluation index to measure the PSNR gap between the enhanced and original compressed sequences on RGB channels.
The comparisons on both CBR mode and CQP mode will be included to fully investigate the performance of compared methods.
We select 200kbps and 800kbps as the typical bit rates for CBR comparison, while select 27 and 37 as the typical QPs for CQP comparison.
%Our method outperforms most of the previous methods in both constant bit rate and constant QP modes.

\begin{table}[htbp]
  \centering
    \caption{Ablation investigation for HIR and ImpFreqUp at QP=37 and BR=800kbps.
The results of $\Delta$PSNR (dB) calculated on the class D is reported. Flops are tested on a $1\times 5\times 96\times 96$ input. }
      %\tabcolsep=0.05cm
  %\resizebox{\linewidth}{!}{
    \begin{tabular}{cccccc}
    \toprule
    \toprule
    \multirow{2}[4]{*}{{ImpFreqUp}} & \multirow{2}[4]{*}{{HIR}} & \multicolumn{2}{c}{{Mode}} & \multirow{2}[4]{*}{{Parameters}} & \multirow{2}[4]{*}{{Flops}} \\
\cmidrule{3-4}          &       & {CBR} & {CQP} &       &  \\
    \midrule
          &       & 0.84  & 0.91  & 5.76M  & 64.93G \\
    \checkmark &       & 1.01  & 1.12  & 5.80M  & 66.52G \\
          & \checkmark & 1.04  & 1.11  & 6.18M  & 65.94G \\
    \checkmark & \checkmark & 1.11  & 1.18  & 6.22M  & 68.13G \\
    \bottomrule
    \bottomrule
    \end{tabular}%
   % }
  \label{tab:ab}%
\end{table}%

\begin{figure*}[htp]
    \centering
     \includegraphics[width=0.9\linewidth]{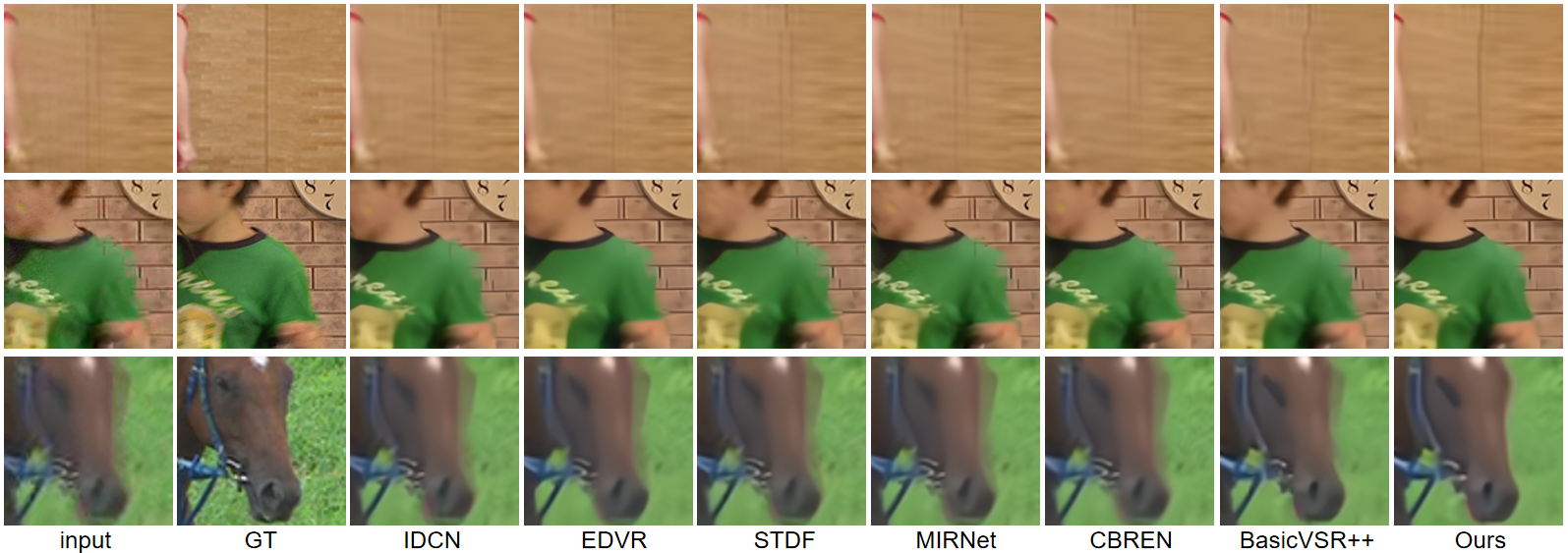}
    \caption{Qualitative results on the state-of-the-art methods and our method on CBR. The test video name (from top to bottom): BasketballPass, PartyScene, and RaceHorses.}
    \label{visualcbr}
\end{figure*}

Table \ref{exp_CBR} and \ref{exp_CQP} present the $\Delta$PSNR in CBR mode and CQP mode. 
The results show that our method performs best in both two modes for average $\Delta$PSNR. Comparing to the CBREN which is specially designed for CBR Videos quality enhancement, our method beat it by 0.07dB and 0.13dB (up to 20\%) in CBR mode. 
Meanwhile, our HFUR also achieve the best performance on all test sequences and beat the second best BasicVSR++ by 0.13dB and 0.17dB in CQP mode.
We also provide the visualized results of compared methods in Fig.~\ref{visualcbr}. 
The compressed patches suffer from various compression artifacts including blocking (in BasketballPass), color bleeding (in PartyScene), and ringing (in RaceHorses). 
Existing methods fail to recognize the artifacts and cannot appropriately suppress the artifacts (e.g., wrong texture on the wall, ringing effect in the horses) or restore the missing details (e.g., border between the clothes and the background). 
Thanks to the powerful frequency-domain information reconstruction, our HFUR could accurately recover the details or textures through the frequency domain and produce more visual pleased results. 
Further experimental and visualized results are available in the \textit{Supplementary Material}.

Moreover, we measure the standard deviation (SD) of frame-level PSNR for each compressed video sequence, to illustrate the quality fluctuation throughout the frames. 
As shown in Tab.~\ref{tab:sd}, our HFUR exhibits the best stability that achieves the smallest quality fluctuation among all compared methods.

\subsection{Ablation Study}

In this section, we examine the effectiveness of each component of the proposed HFUR on both CBR and CQP modes. Since the four videos in class D are representative of the HEVC standard test sequences, we serve as a test set for the ablation experiments and evaluate $\Delta$PSNR at the RGB level. The results are shown in Tab.~\ref{tab:ab}. %As shown in Tab.~\ref{tab:ab}, we adopt the CBREN framework as our baseline (Model 5).
%In model 4, we redesigned the DRM module in CBREN which able to achieve similar performance with half the number of parameters of CBREN.
%\subsubsection{Ablation on the components of DRM}

%In model 4, we redesigned the DRM module in CBREN which able to achieve similar performance with half the number of parameters of CBREN.

\begin{figure}[]
    \centering
     \includegraphics[width=0.7\linewidth]{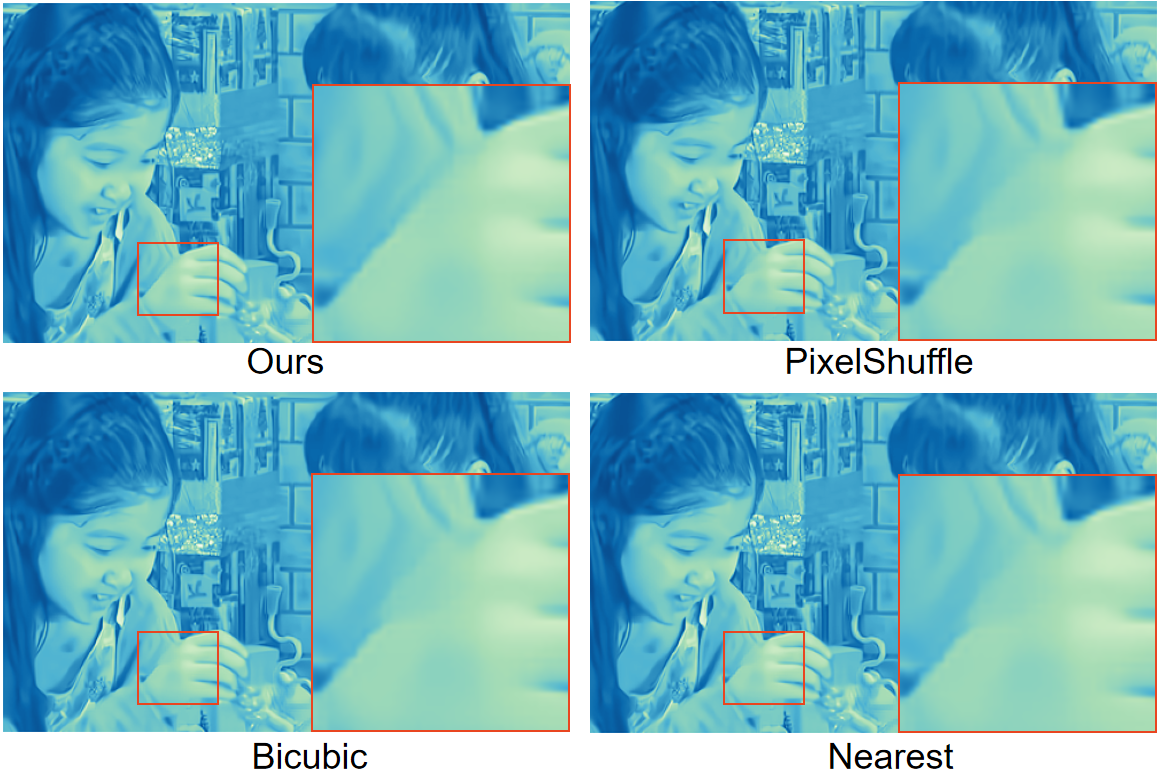}
    \caption{Visual comparison of our ImpFreqUp with other upsampling methods.}
    \label{IFU}
\end{figure}

\textbf{Implicit Frequency Upsampling}.
To demonstrate the effectiveness of our Implicit Frequency Upsampling, we introduce a variant (Row 2) , which adds the ImpFreqUp to base model (Row 1) and improves 0.17dB, 0.21dB at CBR and CQP mode, respectively. Such improvements can be attributed to the fact that our method is able to utilize the frequency domain prior during the upsampling process, preserving more high-frequency information than pixel-domain based upsampling methods.
Additionally, we compare the performance between Pixel shuffle, nearest, bicubic and our ImpFreqUp. The results are presented in Tab.~\ref{tab:IFU}. Quantitative results show that our method exceeds the traditional spatial up-sampling method, and PSNR improves in both CQP and CBR modes. The visualization example in Fig.~\ref{IFU} also shows that our method is capable of better suppressing compression artifacts and provides superior reconstruction of details.

\begin{table}[htbp]
  \centering
  \caption{Ablation study of our upsampling strategy at QP=37 and BR=800.
Experiments are shown with $\Delta$PSNR on D test sequence.}
    \renewcommand\arraystretch{1}
    \begin{tabular}{ccccc}
    \toprule
    \toprule
    
     & {PixelShuffle} & {Nearest} & Bicubic & Proposed\\
    \midrule
    CBR  & 1.04  & 1.01 & 1.02 & \textbf{1.11}\\
    CQP  & 1.11  & 1.05 & 1.08 & \textbf{1.18}\\
    \bottomrule
    \bottomrule
    \end{tabular}%
  \label{tab:IFU}%
\end{table}%

\begin{figure}[]
    \centering
     \includegraphics[width=0.7\linewidth]{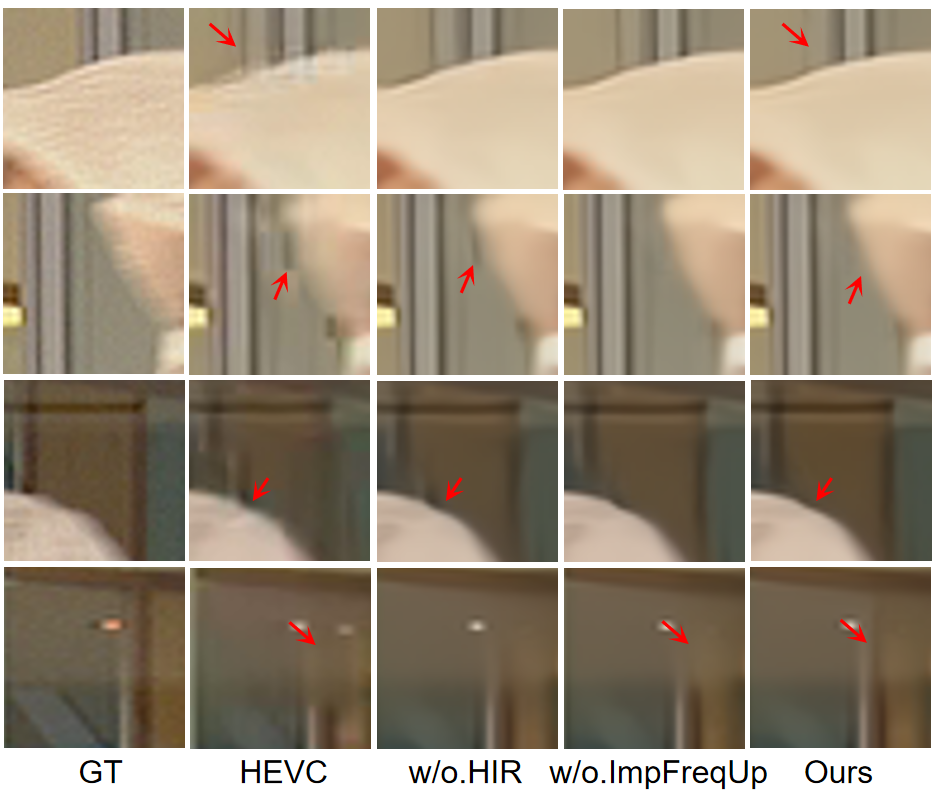}
    \caption{The visual examples for illustrating the effectiveness of our HIR and ImpFreqUp. 
The ImpFreqUp enhances the clarity of edges and details, while the HIR aids in mitigating unnatural artifacts.}
    \label{FIRM_visual}
\end{figure}

\textbf{Hierarchical and iterative refinement module}.
We introduce a variant (Row 3) by inserting a hierarchical and iterative refinement module, which is 0.20 dB higher than the base model (Row 1) in CBR and CQP modes.
This improvement is credited to the alternating iterations of the HIR, leveraging cross-collaboration and information compensation between scales to further refine the features. As shown in Fig.~\ref{FIRM_visual}, introducing HIR could eliminate some unnatural artifacts and promotes the visual quality of the final output.

\section{Conclusion}

In this work, we propose a DNN-based architecture namely HFUR to hierarchically reconstruct frequency information via frequency-based upsampling and iterative feature refinement for effective compressed video quality enhancement. The ImpFreqUp focuses on the propagation of high-frequency information during cross-scalse transfer by leveraging DCT-domain prior via implicit computation. The HIR is used to further refine the feature maps through cross-collaboration between scales and compensation the information.
Extensive experiments show that HFUR achieves the superior performance over the state-of-the-art methods.

%Bibliography
\bibliographystyle{unsrt}  
\bibliography{references}

\end{document}